\documentstyle[prd,aps,preprint]{revtex}

\input{epsf}

\begin{document}
\setcounter{page}{0}
\draft
\tighten
\title{Two-point function of strangeness-carrying vector-currents\\
in two-loop Chiral Perturbation Theory}
\author{Stephan D\"urr\thanks{stephan.duerr@psi.ch}${}^{a,b}$ and
Joachim Kambor\thanks{kambor@physik.unizh.ch}${}^c$}
\address{${}^a$University of Washington, Physics Department,
Seattle, WA 98195-1560, U.S.A.\\
${}^b$Paul Scherrer Institute, Particle Theory Group, 5232 Villigen,
Switzerland\\
${}^c$University of Z\"urich, Institute for Theoretical Physics,
8057 Z\"urich, Switzerland\\
$\mbox{$\qquad$}$}
\vspace{3cm}
\maketitle
\begin{abstract}
We calculate the correlator between two external vector-currents having the
quantum-numbers of a charged kaon. We give the renormalized expression to two
loops in standard chiral perturbation theory in the isospin limit, which, as a
physical result, is finite and scale-independent. Applications include a 
low energy theorem, valid at two loop order, of a flavor breaking combination
of vector current correlators as well as a determination of the
phenomenologically relevant finite $O(p^6)$-counterterm combination $Q_V$
by means of inverse moment finite energy sum rules. This determination 
is less sensitive to isospin-breaking effects than previous attempts.
\end{abstract}
\vfill

Preprint numbers: UW-PT/99-14, ZU-TH/15-98, hep-ph/9907539


\clearpage


\def\J#1#2#3#4{{#1} {\bf #2} (#4), #3}
\def\NPB{{\em Nucl. Phys.} {\bf B}}
\def\NPP{\em Nucl. Phys. (Proc. Suppl.)}
\def\PLB{{\em Phys. Lett.} {\bf B}}
\def\PRD{{\em Phys. Rev.} {\bf D}}
\def\PRL{\em Phys. Rev. Lett.}
\def\CMP{\em Commun. Math. Phys.}
\def\AOP{\em Ann. Phys.}
\def\JPA{{\em J. Phys.} {\bf A}}
\def\JPQ{\em J. Physique (France)}
\def\HPA{\em Helv. Phys. Acta}
\def\ZPC{{\em Z. Phys.} {\bf C}}
\def\JHEP{\em J. High Energy Phys.}

\newcommand{\pad}{\partial}
\newcommand{\pas}{\partial\!\!\!/}
\newcommand{\Dsl}{D\!\!\!\!/\,}
\newcommand{\Psl}{P\!\!\!\!/\;\!}
\newcommand{\Nf}{N_{\!f}}
\newcommand{\hqu}{\hbar}
\newcommand{\ovr}{\over} 
\newcommand{\hal}{{1\ovr2}}
\newcommand{\til}{\tilde}
\newcommand{\pri}{^\prime}
\renewcommand{\dag}{^\dagger}
\newcommand{\<}{\langle}
\renewcommand{\>}{\rangle}
\newcommand{\gaf}{\gamma_5}
\newcommand{\lap}{\triangle}
\newcommand{\trc}{{\rm tr}}
\newcommand{\al}{\alpha}
\newcommand{\be}{\beta}
\newcommand{\ga}{\gamma}
\newcommand{\de}{\delta}
\newcommand{\ep}{\epsilon}
\newcommand{\ve}{\varepsilon}
\newcommand{\ze}{\zeta}
\newcommand{\et}{\eta}
\newcommand{\th}{\theta}
\newcommand{\vt}{\vartheta}
\newcommand{\io}{\iota}
\newcommand{\ka}{\kappa}
\newcommand{\la}{\lambda}
\newcommand{\rh}{\rho}
\renewcommand{\vr}{\varrho}
\newcommand{\si}{\sigma}
\newcommand{\ta}{\tau}
\newcommand{\ph}{\phi}
\newcommand{\vp}{\varphi}
\newcommand{\ch}{\chi}
\newcommand{\ps}{\psi}
\newcommand{\om}{\omega}
\newcommand{\psb}{\overline{\psi}}
\newcommand{\etb}{\overline{\eta}}
\newcommand{\psd}{\psi^{\dagger}}
\newcommand{\chd}{\chi^{\dagger}}
\newcommand{\etd}{\eta^{\dagger}}
\newcommand{\etp}{\eta^{\prime}}
\newcommand{\rch}{{\rm ch}}
\newcommand{\rsh}{{\rm sh}}
\newcommand{\lab}{\overline{\la}}
\renewcommand{\i}{{\rm i}}
\newcommand{\qal}{q_\al}
\newcommand{\qbe}{q_\be}
\newcommand{\qmu}{q_\mu}
\newcommand{\qnu}{q_\nu}
\newcommand{\gmunu}{g_{\mu\nu}}
\newcommand{\Afin}{A_{{\rm fin}}}
\newcommand{\Bbar}{\overline B}
\newcommand{\Jbar}{\overline J}
\newcommand{\Bmu}{B_\mu}
\newcommand{\Bnu}{B_\nu}
\newcommand{\Bmunu}{B_{\mu\nu}}
\newcommand{\Talmu}{T_{\al\mu}}
\newcommand{\Talnu}{T_{\al\nu}}
\newcommand{\Tmunu}{T_{\mu\nu}}
\newcommand{\Cmu}{C_\mu}
\newcommand{\Cnu}{C_\nu}
\newcommand{\Cmunu}{C_{\mu\nu}}
\newcommand{\Umunu}{U_{\mu\nu}}
\newcommand{\Fsq}{F_0^2}
\newcommand{\psq}{p^2}
\newcommand{\qsq}{q^2}
\newcommand{\msq}{m^2}
\newcommand{\mo}{m_1^2}
\newcommand{\mt}{m_2^2}
\newcommand{\Mo}{M_1^2}
\newcommand{\Mt}{M_2^2}
\newcommand{\Met}{M_\et}
\newcommand{\Mka}{M_K}
\newcommand{\Mpi}{M_\pi}
\newcommand{\Mesq}{M_\et^2}
\newcommand{\Mksq}{M_K^2}
\newcommand{\Mpsq}{M_\pi^2}
\newcommand{\mesq}{M_\et^2}
\newcommand{\mksq}{M_K^2}
\newcommand{\mpsq}{M_\pi^2}
\newcommand{\beq}{\begin{equation}}
\newcommand{\eeq}{\end{equation}}
\newcommand{\bdm}{\begin{displaymath}}
\newcommand{\edm}{\end{displaymath}}
\newcommand{\bea}{\begin{eqnarray}}
\newcommand{\eea}{\end{eqnarray}}


\section{Introduction}


The appropriate tool to analyze hadronic processes at very low energies is
chiral perturbation theory (XPT), the low energy effective theory of QCD.
The general structure of XPT is well understood \cite{Wei79,GL84,GL85,Leu94},
and it has been successfully applied at the one- and two-loop level to a vast
body of low-energy data.
Recently, the non-anomalous counterterm Lagrangian at $O(p^6)$ has been
constructed in full generality for chiral $SU(N)$ \cite{BCE99}, thereby closing
a gap left behind in an earlier attempt \cite{FS96}.
The divergent part of the generating functional at the two-loop level has also
been calculated in closed form and is now available \cite{BCEgenfun}.

Despite this advanced state on the theoretical side, phenomenological
applications at the two-loop level are relatively scarce 
\cite{2loopcalcs,GoKa95,BiCoEcGaSa}.
Two reasons may be responsible for
this shortcoming: being an effective theory, XPT introduces new, unknown
coupling constants at each order of the perturbation series. The number of
such unknown parameters is rather large at $O(p^6)$ \cite{BCE99}.
Consequently, it is a priori not clear whether at this order independent
physical processes can be related in a parameter free manner -- at one-loop one
of the main virtues of XPT. The second, more technical reason, is the large
computational effort demanded to perform complete two-loop calculations,
comprising the finite contributions to loop-integrals and counterterms,
in particular in the 3-flavor case.

In this paper we calculate the two-point function of strangeness carrying
vector currents to $O(p^6)$ in standard XPT. Although completely analogous to
the calculation of the correlator of flavor diagonal currents considered in
\cite{GoKa95}, the non-equal masses propagating in the loops make the
present investigation technically more demanding. We use a method
where one-loop subgraphs are renormalized before the second loop-integration
is performed, which helps simplifying intermediate expressions.
Combining our result with the isospin and hypercharge component of the vector
correlator obtained in \cite{GoKa95}, we can form a ``low energy theorem'',
i.e. a {\em parameter free\/} relation between physical quantities,
valid at $O(p^6)$.
This shows that, despite the many coupling constants present at this order,
there are still specific flavor breaking combinations of observables which
can be predicted in terms of meson masses and decay constants only.

A more phenomenological motivation of the present calculation derives from
experimental efforts to determine the spectral functions of strangeness
carrying currents. The ALEPH collaboration has recently published their
analysis of $\tau$-decays into hadronic final states with strangeness
\cite{ALEPHrhos}.
This concatenation of data allows to connect, via inverse moment finite energy
sum rules (IMFESR), the vector current correlator at low energies calculated in
this paper to the asymptotic behavior of the correlator obtained by the
operator product expansion \cite{Na}.
As a result, we are able to fix a coupling constant of the $O(p^6)$
chiral Lagrangian, $Q_V$, from the difference of isovector and strange
component of the vector spectral functions. The constant $Q_V$ has been
estimated in Ref.~\cite{GoKa96} from the difference of isovector and
hypercharge components of the vector spectral functions. However, as pointed
out in \cite{MW98}, that determination is sensitive to isospin violating
corrections affecting mainly the hypercharge current. The strange current on
the other hand is not sensitive to isospin violation, and the determination
of $Q_V$ presented in this paper is therefore more reliable.

The article is organized as follows. Section II contains an outline of the
general strategy employed in the calculation and a derivation of the strange
formfactor which plays a central role in it. In Section III the two-point
correlator $\<T\{V_sV_s\dag\}\>$ between two strangeness carrying external
vector-currents is derived. The physical content of this result is discussed
in Section IV where we also point out several consistency checks. Combining
strange, isospin and hypercharge component of the vector two-point function 
yields a new low energy theorem valid at $O(p^6)$. In section V we present a
determination of the low energy constant $Q_V$ by employing an inverse moment
finite energy sum rule for the difference of isovector and strange component
of the vector current spectral functions. Finally, section VI contains our 
conclusions. Some technical details are collected in the Appendix. 


\section{Form factor calculation and description of the method}


For explicit calculations we shall use the Lagrangian of standard XPT
\beq
{\cal L}={\cal L}^{(2)}+{\cal L}^{(4)}+{\cal L}^{(6)}
\label{lagrangian}
\eeq
where ${\cal L}^{(i)}$ denotes the Lagrangian of order $p^i$. The first two
terms in this expansion can be found in the work by Gasser and Leutwyler
\cite{GL85,Leu94} which also discusses how the meson fields couple to external
sources. The last term, i.e. the one of order $p^6$, has recently been
constructed for the non-anomalous sector in \cite{BCE99}, thereby reducing the
older form given in \cite{FS96} to a minimal set. We shall employ the SU(3)
version of the Lagrangian given in \cite{BCE99}.

\bigskip

From (\ref{lagrangian}), the correlation function between two external vector
currents with the quantum numbers of a kaon could be computed following the
usual procedure: Draw all Feynman diagrams contributing at $O(p^4)$ and
$O(p^6)$ in standard XPT \cite{GL84,GL85} (see Fig.~\ref{Figvv1loop} and
Fig.~\ref{Figvv2loop}), evaluate them and have the divergences in
the loop- and counterterm-contributions canceled to get the finite physical
result valid in $4$ dimensions.
Naively following this recipe would amount to compute the two-loop integrals
in terms of unrenormalized masses and couplings. As it turns out, the two-loop
contributions are significantly shorter if they are expressed in terms of
one-loop renormalized expressions.

\bigskip

A well known method to do this is simply combining diagrams (e) and
(f) in Fig.~\ref{Figvv2loop} into a new single diagram, equivalent in topology
to the one-loop unitary diagram, employing mass- and
wavefunction-renormalization correct to $O(p^4)$.
Likewise, diagrams (b) and (c) can be replaced by a single new diagram,
topologically equivalent to the one-loop tadpole-diagram, without altering
the expression at $O(p^6)$, if one-loop masses are used.
In the remaining six diagrams in Fig.~\ref{Figvv2loop}, the $K,\pi,\et$-masses
may be taken as the one-loop renormalized masses as well, since the
corresponding change shows up at $O(p^8)$ only.

An alternative method uses the strategy of renormalizing intermediate results
in an even more direct, physical way.
It is known as the ``master method'' and was introduced in the context of
XPT in ref.~\cite{BiCoEcGaSa}.
The idea is to start from the one-loop
expression for the formfactor of the strangeness-carrying vector-current with
any of the possible intermediate particle-combinations of the $VV$-correlator
in the final state. 
The correlator follows by contracting this formfactor with its hermitian
conjugate (modulo a modification necessary to avoid double-counting, c.f.
Sec.~III), integrating over internal momenta, adding tadpole- and
counterterm-contributions and completing the remaining renormalization
procedure.

We have calculated $\<T\{V_sV_s\dag\}\>$ following either one of these two
methods and have found the results to agree. For this reason we restrict
ourselves to presenting the calculation in the second, more physical framework.
Below we summarize the results for the strangeness-carrying formfactor, while
section III contains the derivation of the correlator.

\bigskip

\begin{figure}[p]
\begin{center}
\epsfxsize=12cm
\epsffile{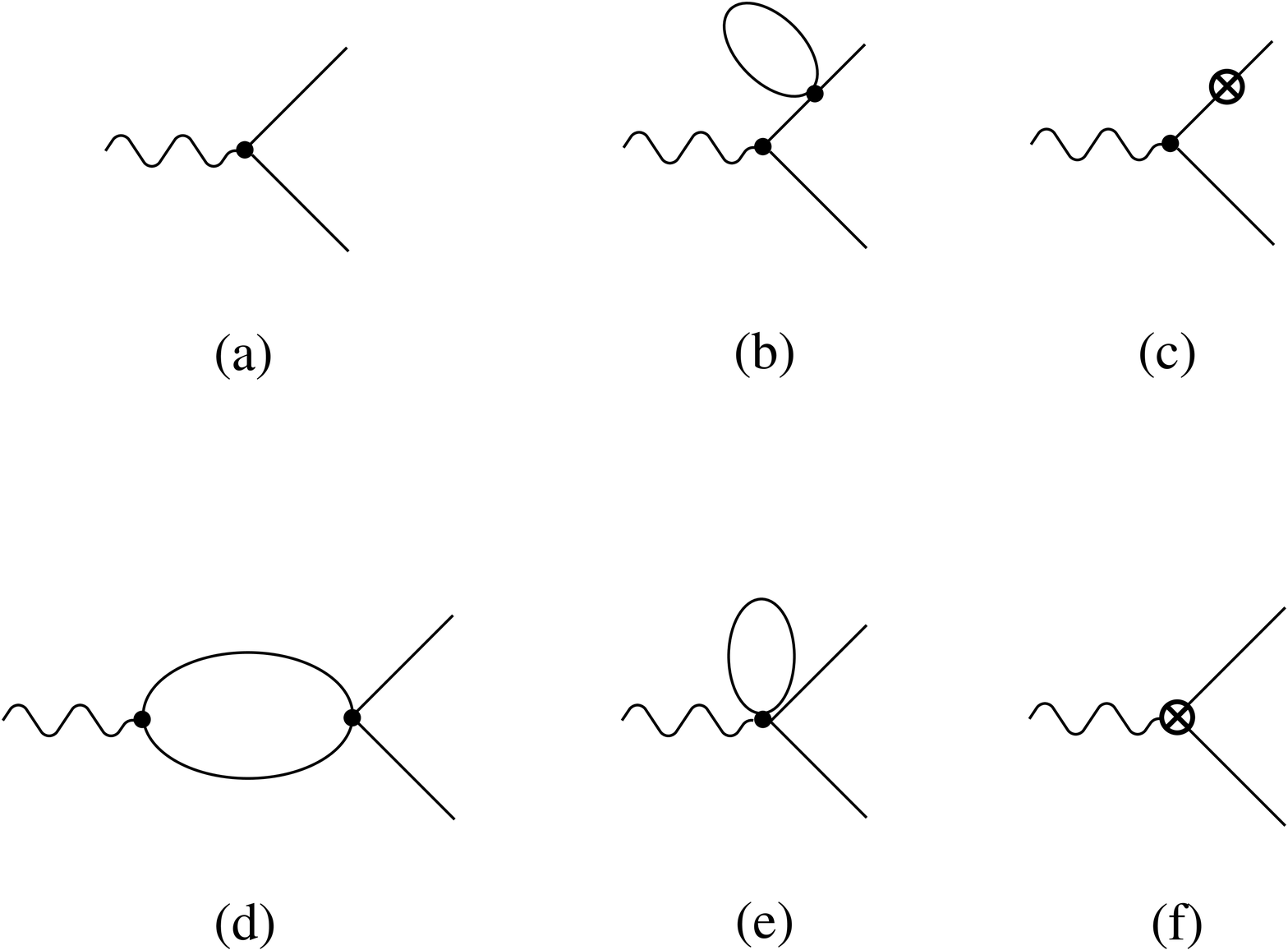}
\vspace{0.5cm}
\caption[]{\sl Feynman graphs for meson form factors at one-loop. The partial
wavefunction renormalization convention used in the calculation affects the
weight of diagrams (b, c) -- see text.}
\label{Figformfactor}
\end{center}
\end{figure}

It is clear that the technical key quantity needed in the second strategy
is the {\em off-shell\/} form-factor $S_\mu[P_1P_2](p_1,p_2)$, i.e. the
amplitude for an external vector-current with momentum $q_\mu=(p_1+p_2)_\mu$
and the quantum numbers of a $K^+$ to decay into any of the combinations
$P_1P_2=(K^+\pi^0,K^+\et,K^0\pi^+)$ with {\em arbitrary\/} momenta $p_1$ and
$p_2$, respectively.
Note that $S_\mu$ is not a physical quantity, i.e. the result depends on the
choice of interpolating fields for the Goldstone bosons. However, the
correlator we are aiming at involves an integration over these fields and is
independent of this choice.
For convenience, the formfactor shall be separated into a symmetric and an
antisymmetric piece w.r.t. the outgoing momenta
\beq
S_\mu[P_1\,P_2]=
S_{+}[P_1\,P_2]\,(p_1+p_2)_\mu+S_{-}[P_1\,P_2]\,(p_1-p_2)_\mu
\label{formfac}
\eeq
and the convention is that $p_1$ denotes the momentum of the strange
meson ($P_1=K^+,K^0$), whereas $p_2$ is the momentum of the
non-strange meson ($P_2=\pi^0,\eta,\pi^+$).
Fig.~\ref{Figformfactor} displays diagrammatically the relevant contributions.
Diagram (a) represents the leading $O(p^2)$-contribution, whereas diagrams
(b,~c) as well as (d,~e,~f) are of order $O(p^4)$.
The latter three are frequently referred to as the ``unitary''-, ``tadpole''-
and ``counterterm''-contributions.
The sum of these six diagrams corresponds to a single formfactor-diagram in
the effective theory
[in the field-theoretic sense, i.e. in the theory where
all quantum corrections have been absorbed into the propagators and vertices
of the effective action and Greens functions are constructed from tree-level
diagrams only]
where the two outgoing mesons carry the full wavefunction-renormalization
factors $Z_{P_1}, Z_{P_2}$, with $Z_P=1-A_P$ and
\bea
A_\pi
&=&
{\i\ovr\Fsq}(-{2\ovr3}A(\Mpsq)-{1\ovr3}A(\Mksq))
+{8\ovr\Fsq}((2\Mksq+\Mpsq)L_4+\Mpsq L_5)
\label{Api}
\\
A_K
&=&
{\i\ovr\Fsq}(-{1\ovr4}A(\Mpsq)-{1\ovr2}A(\Mksq)-{1\ovr4}A(\Mesq))
+{8\ovr\Fsq}((2\Mksq+\Mpsq)L_4+\Mksq L_5)
\label{Aka}
\\
A_\eta
&=&
{\i\ovr\Fsq}(-A(\Mksq))
+{8\ovr\Fsq}((2\Mksq+\Mpsq)L_4+\Mesq L_5)
\label{Aet}
\;.
\eea
However, in the light of the application to follow, it is extremely convenient
to have the external legs carry a factor $\sqrt{Z_P}$, rather than $Z_P$ or 1.
Since $Z_P=1-A_P$ implies $\sqrt{Z_P}=1-\hal A_P+O(p^4)$ and since we
are interested in $S_\mu$ to $O(p^4)$, this ``$\sqrt{Z_P}$-convention''
amounts to defining $S_\mu[P_1P_2]$ as a weighted sum:
Diagrams (a,~d,~e,~f) are included with their full weight, but diagrams
(b,~c) with a factor $\hal$ only.
Note that renormalizability of the on-shell formfactor is not affected
by such a reweighting as long as diagrams (b) and (c) receive the same factor
and the group (d,~e,~f) also receives a common factor.
In practice, on-shell renormalizability is implemented as follows:
First replace $p_1^2\rightarrow \Mksq,\;p_2^2\rightarrow M_{P_2}^2$
($P_2=\pi^+,\pi^0,\et$), where $M_P$ denotes the {\em one-loop\/} mass of
particle $P$.
Next, expand all integrals in $1/(d-4)$, or equivalently in $\lab$, as is
explained in the Appendix, and simultaneously the renormalization constants
according to
\beq
L_i\rightarrow\mu^{d-4}\cdot
(L_i^{(1)}\lab+L_i^{(0)}+O(\lab^{\,-1}))
\;.
\label{Lexp}
\eeq
Finally, check that there is indeed a set of numerical constants
$\{L_i^{(1)}\}$ for which all contributions in $\lab$ cancel exactly.
These ``divergent'' parts of the one-loop counterterms agree of course
with their standard values  $L_i^{(1)}=\Gamma_i$ \cite{GL84,GL85}.

In summary, the strange {\em off-shell\/} formfactor takes the form
(\ref{formfac}) with
\bea
S_{+}[K^+\pi^0]={1\ovr\Fsq}\;
&\Big\{&
-{1\ovr16}[-5\Afin(\mpsq)+2\Afin(\mksq)+3\Afin(\mesq)]
\nonumber
\\
&{}&
-2\i\,L_5^{(0)}(\mksq-\mpsq)+\i\,L_9^{(0)}(p_1^2-p_2^2)
\nonumber
\\
&{}&
+{1\ovr48}[\mksq+\mpsq+5(p_1^2+p_2^2-3\qsq)]\,
{\mksq-\mpsq\ovr\qsq}\,\Bbar(\qsq,\mksq,\mpsq)
\nonumber
\\
&{}&
-{1\ovr48}[3\mksq+3\mpsq+3(p_1^2+p_2^2-3\qsq)]\,
{\mksq-\mesq\ovr\qsq}\,\Bbar(\qsq,\mksq,\mesq)
\nonumber
\\
&{}&
+{3\ovr16}\,{p_1^2-p_2^2\ovr\qsq}
[(T^{(0)}_{\rm fin}-T^{(1)}_{\rm fin})(\qsq,\mksq,\mpsq)+
(T^{(0)}_{\rm fin}-T^{(1)}_{\rm fin})(\qsq,\mksq,\mesq)]\;
\Big\}
\!\!
\label{S+pnu}
\\
S_{+}[K^+\et]={\sqrt{3}\ovr\Fsq}\;
&\Big\{&
-{3\ovr16}[\Afin(\mpsq)-2\Afin(\mksq)+\Afin(\mesq)]
\nonumber
\\
&{}&
-2\i\,L_5^{(0)}(\mksq-\mesq)+\i\,L_9^{(0)}(p_1^2-p_2^2)
\nonumber
\\
&{}&
+{1\ovr48}[\mksq-7\mpsq-3(p_1^2+p_2^2-3\qsq)]\,
{\mksq-\mpsq\ovr\qsq}\,\Bbar(\qsq,\mksq,\mpsq)
\nonumber
\\
&{}&
-{1\ovr48}[19\mksq-5\mpsq+3(p_1^2+p_2^2-3\qsq)]\,
{\mksq-\mesq\ovr\qsq}\,\Bbar(\qsq,\mksq,\mesq)
\nonumber
\\
&{}&
+{3\ovr16}\,{p_1^2-p_2^2\ovr\qsq}
[(T^{(0)}_{\rm fin}-T^{(1)}_{\rm fin})(\qsq,\mksq,\mpsq)+
(T^{(0)}_{\rm fin}-T^{(1)}_{\rm fin})(\qsq,\mksq,\mesq)]\;
\Big\}
\!\!
\label{S+eta}
\\
S_{+}[K^0\pi^+]=\sqrt{2}\;&S&{}_q[K^+\pi^0]
\label{S+ppl}
\eea
\bea
S_{-}[K^+\pi^0]
&=&
-{\i\ovr2}+{1\ovr\Fsq}\;
\Big\{
-{3\ovr16}[\Afin(\mpsq)+2\Afin(\mksq)+\Afin(\mesq)]
-\i L_9^{(0)}\qsq
\nonumber
\\
&{}&
\qquad\qquad\quad\;
+{3\ovr16}\,
[T^{(1)}_{\rm fin}(\qsq,\mksq,\mpsq)+T^{(1)}_{\rm fin}(\qsq,\mksq,\mesq)]\;
\Big\}
\label{S-pnu}
\\
S_{-}[K^+\et]
&=&
\sqrt{3}\;S_{-}[K^+\pi^0]
\label{S-eta}
\\
S_{-}[K^0\pi^+]
&=&
\sqrt{2}\;S_{-}[K^+\pi^0]
\quad,
\label{S-ppl}
\eea
where the particle masses $M_P$ refer to the {\em one-loop renormalized
masses\/} and the integrals are all finite and explicitly given in the Appendix.

It is worth mentioning that this result reduces to the expression known in the
literature, if we restrict ourselves to {\em on-shell\/} outgoing momenta.
Employing the notation of \cite{GL85} 
\beq
\<K^+,p'|\sqrt{2}\;\overline u\gamma_\mu s| \pi^0,p\>
=
(p'_\mu+p_\mu)f^{K\pi}_+(t)+(p'_\mu-p_\mu)f^{K\pi}_-(t)
\label{formfaconshell}
\eeq
and
\beq
f^{P_1P_2}_0(t)\equiv f^{P_1P_2}_+(t)+{t\ovr\Mo-\Mt}\;f^{P_1P_2}_-(t)
\;,
\qquad
t=(p'_\mu-p_\mu)^2
\;,
\label{fnotdecomp}
\eeq
our result simplifies in the on-shell limit to 
(\ref{formfaconshell},~\ref{fnotdecomp}) with
\bea
f^{K\pi}_+(\qsq) = f^{K\eta}_+(\qsq) &{}&
\nonumber
\\
= 1+{\i\ovr \Fsq}\;
&\Big\{&
{3\ovr8}\;[T^{(1)}_{\rm fin}(\qsq,\mksq,\mpsq)-\Afin(\mksq)-\Afin(\mpsq)]-
\i \qsq L_9^{(0)}+(\pi\rightarrow \eta)
\Big\}
\label{f+kapi}
\\
f^{K\pi}_0(\qsq)=1+{1\ovr\Fsq}\;
&\Big\{&
{\i\ovr8}\;{\qsq\ovr\Mksq-\Mpsq}\;
[5\Afin(\mpsq)-2\Afin(\mksq)-3\Afin(\mesq)]
+4\,L_5^{(0)}\qsq
\label{fnotkapi}
\\
&{}&
+{1\ovr8}\;[5\qsq-2(\mksq+\mpsq)-{3(\mksq-\mpsq)^2\ovr\qsq}]\;
{1\ovr\i} \Bbar(\qsq,\mksq,\mpsq)
\nonumber
\\
&{}&
+{1\ovr24}\;[3\qsq-2(\mksq+\mpsq)-{(\mksq-\mpsq)^2\ovr\qsq}]\;
{1\ovr\i} \Bbar(\qsq,\mksq,\mesq)
\;\Big\}
\nonumber
\\
f^{K\et}_0(\qsq)=1+{1\ovr\Fsq}\;
&\Big\{&
-{3\i\ovr8}\;{\qsq\ovr\Mksq-\Mesq}\;
[\Afin(\mpsq)-2\Afin(\mksq)+\Afin(\mesq)]
+4\,L_5^{(0)}\qsq
\label{fnotkaet}
\\
&{}&
+{3\ovr8}\;[3\qsq-2(\mksq+\mpsq)-{(\mksq-\mpsq)^2\ovr\qsq}]\;
{1\ovr\i} \Bbar(\qsq,\mksq,\mpsq)
\nonumber
\\
&{}&
+{1\ovr24}\;[-9\qsq+2(\mksq+9\mesq)-{9(\mksq-\mesq)^2\ovr\qsq}]\;
{1\ovr\i} \Bbar(\qsq,\mksq,\mesq)
\;\Big\}
\;.
\nonumber
\eea
This coincides with the result given in \cite{GL85}.


\section{Two-point Function Calculation}


Here we give the derivation of the technical key quantity of this paper, the
correlator $\<T\{V_sV_s\dag\}\>$ between two external vector currents with
the quantum numbers of a $K^\pm$, respectively.
The convention is that $V_s$ creates a multi-particle state with the
quantum numbers of a $K^+$, or annihilates a state with the quantum numbers
of a $K^-$, i.e.
\beq
V_s^\mu=\bar q\ga^\mu{\la_4+\i\la_5\ovr2\sqrt{2}}q
\label{Vsdef}
\eeq
with $\la_a$ the standard Gell-Mann matrices and $q=(u,d,s)$.
Defining the momentum-space representation of the correlator by
\beq
\Pi_{V,s}(\qmu,\qnu)=
\i\int d^4x\;e^{\i q\cdot x}\<0|T\{V_s^\mu(x)\;V_s^\nu(0)\dag\}|0\>
\; ,
\label{Pisdef}
\eeq
and decomposing the latter according to
\beq
\Pi_{V,s}(\qmu,\qnu)=
(\qmu\qnu-\qsq\gmunu)\;\Pi^{(1)}_{V,s}(\qsq)+
\qmu\qnu\;\Pi^{(0)}_{V,s}(\qsq)
\; ,
\label{Sdecomp}
\eeq
the result will be given in terms of $\Pi^{(1)}_{V,s}$ and $\Pi^{(0)}_{V,s}$.


\subsection{Calculation}

\begin{figure}[p]
\begin{center}
\epsfxsize=15cm
\epsffile{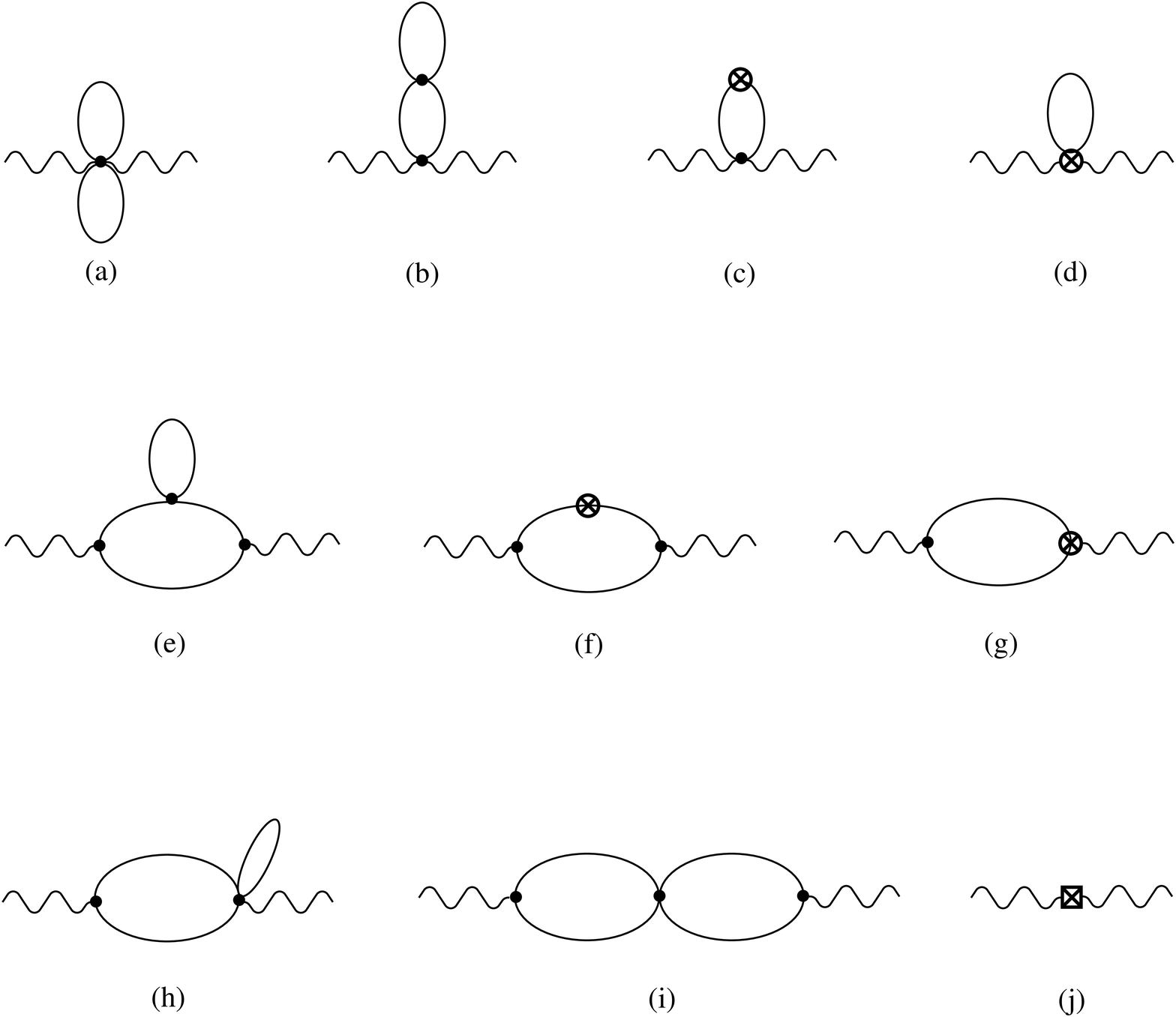}
\vspace{0.5cm}
\caption[]{\sl Two-loop Feynman graphs for $\Pi_{V,s}$. In addition to the
conventions as in Fig.~1, the crossed square denotes a vertex from the
lagrangian of order $p^6$. Diagrams (g) and (h) shall comprise the
corresponding modification to the incoming vector-vertex as well.}
\label{Figvv2loop}
\end{center}
\end{figure}

For convenience we shall consider the ``unitary'', the ``tadpole'' and the
``counterterm'' contributions to $\i\Pi_{V,s}$ separately, i.e. the latter
is decomposed as
\beq
\i\Pi_{V,s}=
\i\Pi^{\rm uni}+\i\Pi^{\rm tad}+\i\Pi^{\rm cnt}
\; ,
\label{PiDec}
\eeq
where each term on the r.h.s. shall contain both the contribution at $O(p^4)$
and at $O(p^6)$ in standard chiral perturbation theory.
In diagrammatic language the two-loop ``unitary'' contribution
$\i\Pi_{p^6}^{\rm uni}$ is given by the sum of diagrams (e,~f,~g,~h,~i) in
Fig.~\ref{Figvv2loop}.
The ``tadpole'' contribution $\i\Pi_{p^6}^{\rm tad}$ denotes the sum of
diagrams (a,~b,~c,~d), and the ``counterterm'' contribution
$\i\Pi_{p^6}^{\rm cnt}$ is just diagram (j) in Fig.~\ref{Figvv2loop}.

The key ingredient in the approach to the correlator $\<T\{V_sV_s\dag\}\>$ we
are describing is the observation that the ``unitary piece''
$\i\Pi_{p^4+p^6}^{\rm uni}$ can be computed from the renormalized strange
formfactor obtained in the previous chapter.
The prescription is: Contract the formfactor $S_\mu[P_1,P_2](p_1,p_2)$
with its hermitian conjugate and sum over all possible intermediate
particle $P_1,P_2$ and momenta $p_1,p_2$.
In practice, several points have to be paid attention to.

First, we should mention why we defined the formfactor such that outgoing 
mesons carry only the square-root of the usual wavefunction renormalization
factor, but neither $Z$ nor $1$: This choice allows us to proceed by splicing
two of these formfactors together without having to worry again about
wavefunction renormalization~-- the internal propagators generated in this
process will automatically carry the full factor $Z$.
The second point is that splicing the formfactor with its hermitian conjugate
will yield the ``eyeglass''-diagram (i) in Fig.~\ref{Figvv2loop} twice instead
of only once. Thus one has to subtract half of the ``unitary'' diagram (d) in
Fig.~\ref{Figformfactor} from the formfactor before doing the splicing with the
resulting object.  In summary, the unitary piece of the correlator --~which
shall include the contributions both at $O(p^4)$ and at $O(p^6)$ in the chiral
expansion~-- is constructed as
\bea
\i\Pi_{p^4+p^6}^{\rm uni}
&=&
\sum_{P_1,P_2}\int{d^d\ell\ovr(2\pi)^4}\;
{\overline{S}_\mu[P_1,P_2]\cdot\overline{S}_\nu[P_1,P_2]\;(\i)^2\ovr
(\ell^2-\Mo)((q-\ell)^2-\Mt)}
\nonumber
\\
&=&
\sum_{P_1,P_2}\int{d^d\ell\ovr(2\pi)^4}\;
\Big\{
{(2\ell-q)_\mu(2\ell-q)_\nu\;
({S_-^{(0)}}^2+2\overline{S}_-^{(2)}\cdot S_-^{(0)}+O(p^4))\;
(\i)^2\ovr
(\ell^2-\Mo)((q-\ell)^2-\Mt)}
\nonumber
\\
&{}&
\qquad\qquad\qquad+
{((2\ell-q)_\mu\qnu+\qmu(2\ell-q)_\nu)\;
(S_-^{(0)}\cdot\overline{S}_+^{(2)}+O(p^4))\;
(\i)^2\ovr
(\ell^2-\Mo)((q-\ell)^2-\Mt)}
\Big\}
\;.
\label{splicing}
\eea
In (\ref{splicing}) $\overline{S}_\pm$ denotes the ``modified formfactor'' in
which the unitary contribution is included with a factor $\hal$, i.e.
$\overline{S}_\pm=S_\pm-\hal S_\pm^{\rm uni}$, which clearly affects only
the $O(p^2)$-contributions in $S_+[P_1,P_2]=S_+^{(2)}[P_1,P_2]+O(p^4)$ and
$S_-[P_1,P_2]=S_-^{(0)}[P_1,P_2]+S_-^{(2)}[P_1,P_2]+O(p^4)$, but not
$S_-^{(0)}[P_1,P_2]$.
Of course $\overline{S}_\pm$ is not renormalizable, i.e. including this factor
$\hal$ into the formfactor already in the previous section is not an option.
Note that this procedure generates an expression for $\i\Pi_{p^4+p^6}^{\rm uni}$
where the $M_P$ refer to the {\em one-loop renormalized masses\/}.

\begin{figure}[p]
\begin{center}
\epsfxsize=12cm
\epsffile{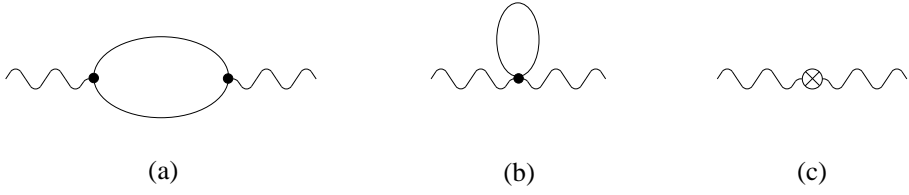}
\vspace{0.5cm}
\caption[]{\sl One-loop Feynman graphs for $\Pi_{V,s}$. Dots and crossed
circles denote vertices from the lagrangian of order $p^2$ and $p^4$
respectively. Wavy lines are external vector currents, straight lines 
denote Goldstone boson propagators.}
\label{Figvv1loop}
\end{center}
\end{figure}

Finally, the ``unitary'' piece (\ref{splicing}) must be augmented with the
``tadpole'' and the ``counterterm'' contribution.
$\i\Pi_{p^4+p^6}^{\rm tad}$ is put together from diagrams (a,~b,~c,~d) in
Fig.~\ref{Figvv2loop} plus the single tadpole-diagram at $O(p^4)$, i.e. diagram
(b) in Fig.~\ref{Figvv1loop}.
$\i\Pi_{p^4+p^6}^{\rm cnt}$ represents the sum of diagram (j) in
Fig.~\ref{Figvv2loop} and the single counterterm-diagram at $O(p^4)$, i.e.
diagram (c) in Fig.~\ref{Figvv1loop}.
It is important to note that both $\i\Pi_{p^4+p^6}^{\rm tad}$ and
$\i\Pi_{p^4+p^6}^{\rm cnt}$ happen to combine a $O(p^4)$ contribution and
a $O(p^6)$ contribution in such a way that the total ``tadpole'' or
``counterterm'' contribution depends on the {\em one-loop renormalized
masses\/} only.


\subsection{Renormalization}

Having added the various contributions to $S_{p^4+p^6}$ the only task left is
renormalizing the result, i.e. rewriting it in such a way that it can be
continued to $d=4$ dimensions.

First we have to remind ourselves that the correlator is constructed from
various elements, some of which have undergone renormalization at $O(p^4)$
already.
In practice, this means that all $L_i^{(0)}$ showing up in this expression
must be replaced $L_i^{(0)}\rightarrow L_i^{(0)}+L_i^{(-1)}\lab^{\,-1}$
before the so-far untouched constants get expanded according to
\bea
L_i&\rightarrow&\mu^{d-4}\cdot
(L_i^{(1)}\lab+L_i^{(0)}+L_i^{(-1)}\lab^{\,-1}+O(\lab^{\,-2}))
\label{Lexpand}
\\
C_j&\rightarrow&\mu^{2d-8}\cdot
(C_j^{(2)}\lab^2+C_j^{(1)}\lab+C_j^{(0)}+O(\lab^{\,-1}))
\label{Bexpand}
\eea
as is necessary in a two-loop calculation.
Next, all integrals must be expanded in $1/(d-4)$, or equivalently
in $\lab$, the latter being defined in (\ref{labdef}) in the Appendix.
For some more standard definitions in this ``$\lab$-scheme'' the reader is
referred to \cite{GoKa95}.
An explicit calculation in this scheme, where technical aspects are presented
with great care, is the analogous computation of the axial correlator to
$O(p^6)$ in chiral perturbation theory \cite{GoKa98a}.
A discussion of the relationship of the ``$\lab$-scheme'' to the
$\overline{\rm MS}$-scheme employed in \cite{BiCoEcGaSa} is also found 
in \cite{GoKa98a} .

The final step is checking that, after the remaining $L_i^{(1)}$ have been
assigned their standard values $\Gamma_i$ \cite{GL84,GL85}, there is indeed
a set of numerical constants $\{C_j^{(2)}, C_j^{(1)}, L_i^{(0)}\}$ for which
all contributions to the correlator with a positive power of $\lab$ cancel
exactly.
Demanding that the total contribution in proportion to $\lab^{\,2}$ is zero
we find the constraints
\bea
0&=&
5-8(4 C_{38}^{(2)}+C_{91}^{(2)})\Fsq
\label{constr1}
\\
0&=&
1+8 C_{93}^{(2)}\Fsq
\label{constr2}
\\
0&=&
C_{61}^{(2)}\Fsq
\label{constr3}
\\
0&=&
C_{62}^{(2)}\Fsq
\label{constr4}
\eea
whereas requiring the total contribution in proportion to $\lab$ to be zero
yields
\bea
0&=&
10L_5^{(0)}-3(4 C_{38}^{(1)}+C_{91}^{(1)})\Fsq
\label{constr5}
\\
0&=&
L_9^{(0)}+C_{93}^{(1)} \Fsq
\label{constr6}
\\
0&=&
3(L_9^{(0)}+L_{10}^{(0)})-4 C_{61}^{(1)}\Fsq
\label{constr7}
\\
0&=&
(L_9^{(0)}+L_{10}^{(0)})-4 C_{62}^{(1)}\Fsq
\;.
\label{constr8}
\eea
Constraints (22-29) prove consistent with what is known in the literature
\cite{GoKa95,BCEgenfun}.


\subsection{Results}

In summary, the renormalized correlator $\<T\{V_sV_s\dag\}\>$ between two
external vector currents carrying the quantum numbers of a $K^\pm$
respectively, is found to take the form (\ref{Pisdef},~\ref{Sdecomp}) with
\bea
\Pi^{(1)}_{V,s}=
&\Big\{&
-2(L_{10}^{(0)}+2H_1^{(0)})
+{3\i\ovr4\qsq}[T_{\rm fin}^{(1)}(\qsq,\mksq,\mpsq)-\Afin(\mksq)-\Afin(\mpsq)]
\nonumber
\\
&{}&
+{3\i\ovr4\qsq}[T_{\rm fin}^{(1)}(\qsq,\mksq,\mesq)-\Afin(\mksq)-\Afin(\mesq)]
\Big\}
\nonumber
\\
+\,\,
&\Big\{&
-2\i L_5^{(0)}{(\mksq-\mpsq)\ovr\qsq}
[3\Afin(\mpsq)-2\Afin(\mksq)-\Afin(\mesq)]
\nonumber
\\
&{}&
+3\i L_9^{(0)}
[T_{\rm fin}^{(1)}(\qsq,\mksq,\mpsq)+T_{\rm fin}^{(1)}(\qsq,\mksq,\mesq)]
\nonumber
\\
&{}&
+3\i L_{10}^{(0)}
[\Afin(\mpsq)+2\Afin(\mksq)+\Afin(\mesq)]
\nonumber
\\
&{}&
-{3\ovr32\qsq}
[-5\Afin(\mpsq)^2+4\Afin(\mpsq)\Afin(\mksq)+6\Afin(\mpsq)\Afin(\Mesq)
\nonumber
\\
&{}&
\qquad\quad\;
+4\Afin(\mksq)^2-12\Afin(\mksq)\Afin(\mesq)+3\Afin(\mesq)^2]
\nonumber
\\
&{}&
-{9\ovr32\qsq}
[T_{\rm fin}^{(1)}(\qsq,\mksq,\mpsq)-\Afin(\mksq)-\Afin(\mpsq)
\nonumber
\\
&{}&
\qquad\quad\;
+T_{\rm fin}^{(1)}(\qsq,\mksq,\mesq)-\Afin(\mksq)-\Afin(\mesq)]^2
\nonumber
\\
&{}&
-{(\mksq-\mpsq)^2\ovr\qsq}O_V-P_V\qsq-4\mksq Q_V-4(2\mksq+\mpsq)R_V
\Big\}\cdot{1\ovr F_0^2}
\label{Sonetot}
\\
\Pi^{(0)}_{V,s}=
&\Big\{&
-{3\i\ovr4}{(\mksq-\mpsq)^2\ovr q^4}\Bbar(\qsq,\mksq,\mpsq)
-{3\i\ovr4}{(\mksq-\mesq)^2\ovr q^4}\Bbar(\qsq,\mksq,\mesq)
\Big\}
\nonumber
\\
+\,\,
&\Big\{&
2\i L_5^{(0)}{(\mksq-\mpsq)\ovr\qsq}
[3\Afin(\mpsq)-2\Afin(\mksq)-\Afin(\mesq)]
\nonumber
\\
&{}&
+{3\ovr32\qsq}
[-5\Afin(\mpsq)^2+4\Afin(\mpsq)\Afin(\mksq)+6\Afin(\mpsq)\Afin(\Mesq)
\nonumber
\\
&{}&
\qquad\quad\;
+4\Afin(\mksq)^2-12\Afin(\mksq)\Afin(\mesq)+3\Afin(\mesq)^2]
\nonumber
\\
&{}&
+\ka_{K\pi}\cdot{\mksq-\mpsq\ovr\qsq}\Bbar(\qsq,\mksq,\mpsq)
+\ka_{K\et}\cdot{\mksq-\mesq\ovr\qsq}\Bbar(\qsq,\mksq,\mesq)
\nonumber
\\
&{}&
+{(\mksq-\mpsq)^2\ovr\qsq}O_V
\Big\}\cdot{1\ovr F_0^2}
\label{Snottot}
\eea
where
\bea
\ka_{K\pi}
&=&
{3\ovr16}[5\Afin(\mpsq)-2\Afin(\mksq)-3\Afin(\mesq)]-6\i(\mksq-\mpsq)L_5^{(0)}
\nonumber
\\
&{}&
+{3\ovr32}[3{(\mksq-\mpsq)^2\ovr\qsq}+2(\mksq+\mpsq)-5\qsq]
{(\mksq-\mpsq)\ovr\qsq}\;\Bbar(\qsq,\mksq,\mpsq)
\nonumber
\\
&{}&
+{3\ovr32}[-9{(\mksq-\mesq)^2\ovr\qsq}-2(\mksq+\mpsq)+3\qsq]
{(\mksq-\mesq)\ovr\qsq}\;\Bbar(\qsq,\mksq,\mesq)
\label{Kkapidef}
\\
\ka_{K\et}
&=&
-{9\ovr16}[\Afin(\mpsq)-2\Afin(\mksq)+\Afin(\mesq)]-6\i(\mksq-\mesq)L_5^{(0)}
\nonumber
\\
&{}&
+{3\ovr32}[-{(\mksq-\mpsq)^2\ovr\qsq}-2(\mksq+\mpsq)+3\qsq]
{(\mksq-\mpsq)\ovr\qsq}\;\Bbar(\qsq,\mksq,\mpsq)
\nonumber
\\
&{}&
+{3\ovr32}[3{(\mksq-\mesq)^2\ovr\qsq}-{2\ovr3}(13\mksq-3\mpsq)+3\qsq]
{(\mksq-\mesq)\ovr\qsq}\;\Bbar(\qsq,\mksq,\mesq)
\;.
\label{Kkaetdef}
\eea
In (\ref{Sonetot}) and (\ref{Snottot}) the contribution at $O(p^4)$ has been
separated from the one at $O(p^6)$ and we have used the abbreviations
\bea
O_V&=&4(4 C_{38}^{(0)}+C_{91}^{(0)}) F_0^2-(40/3)L_5^{(-1)}
\label{OVdef}
\\
P_V&=&4 C_{93}^{(0)} F_0^2+4L_9^{(-1)}
\label{PVdef}
\\
Q_V&=&4 C_{61}^{(0)} F_0^2-3(L_9^{(-1)}+L_{10}^{(-1)})
\label{QVdef}
\\
R_V&=&4 C_{62}^{(0)} F_0^2-(L_9^{(-1)}+L_{10}^{(-1)})
\label{RVdef}
\eea
as these combinations of (finite parts of) $O(p^6)$-counterterms occur quite
generally \cite{GoKa95}.

While completing this paper, we received a preprint \cite{AmBiTa} where this
result has been obtained independently.
Up to an overall factor $2$ (due to a factor $\sqrt{2}$ difference in the
definition of the currents) the results agree.


\section{Discussion and low energy theorem}


In Sect. III we have obtained the fully renormalized finite two-loop 
expressions for the correlator of strangeness carrying vector currents. 
Here we shall first discuss the general structure of the result and point 
out several consistency checks which have to be satisfied on general grounds.
Then we combine our result with the corresponding expressions for 
the isospin and hypercharge currents \cite{GoKa95} to obtain a low energy 
theorem valid at $O(p^6)$. 


\subsection{Consistency checks}
 
The correlators $\Pi_{V,s}^J$ given in Eqns.~(\ref{Sonetot},~\ref{Snottot})
involve the finite functions $\Afin$, $\Bbar$ defined in the Appendix as well
as the renormalized coupling constants $L_{5,9,10}^{(0)}, H_1^{(0)}, O_V, P_V,
Q_V, R_V$. Note that upon taking the (1+0) component, $O_V$ and the direct
dependence on $L_5^{(0)}$ drop out (the indirect, through $\ka$, remains).
Our result satisfies the following consistency checks:
\begin{itemize}
\item
The low energy constants and the function $\Afin(M_P^2)$ depend on the
renormalization scale $\mu$. Upon using Eqns.~(\ref{constr1}-\ref{constr8})
as well as the renormalization group equations for the pertinent low 
energy constants \cite{GoKa95} it can be shown that the sum of all
contributions is scale independent.
\item
The SU(3) limit of equal quark masses provides another check: The  $J=0$
component vanishes identically, as it is in proportion to the difference of
two meson masses squared. The $J=1$ component reduces to the known results for
the isospin or hypercharge component in the degenerate case. Although
non-trivial, this consistency check concerns the SU(3) conserving part of the
amplitude only.
\item
Going away from $SU(3)$, the Ademollo Gatto theorem \cite{AdGaBeSi} is
satisfied: The $SU(3)$ breaking effect appears only in the second order in
the quark masses, i.e. $\propto(m_s-\hat m)^2$.
\item
Individual spin components $J=1,0$ exhibit poles at $q^2=0$. These poles must
be of kinematical origin since there is no single particle state present which
could generate such a singularity. Using --~instead of (\ref{Sdecomp})~-- the
new decomposition
\beq
\Pi_{V,s}(\qmu,\qnu)=
(\qmu\qnu-\qsq\gmunu)\;\Pi^{(1+0)}_{V,s}(\qsq)+
\gmunu\;q^2\Pi^{(0)}_{V,s}(\qsq)
\; ,
\label{Pidecompnew}
\eeq
both new components, $\Pi^{(1+0)}_{V,s}(\qsq)$ and  $q^2 \Pi^{(0)}_{V,s}(\qsq)$
are found to be regular at $q^2=0$. It is thus more convenient to work with
these combinations, in particular when considering dispersive representations
based on analyticity assumptions. 
\item
The explicit form (\ref{Sonetot},~\ref{Snottot}) of the vector correlator 
allows us to extract the corresponding spectral functions 
$\rho_{V,s}^{(J)} \equiv {1 \ovr \pi} {\rm Im} \Pi_{V,s}^{(J)}$ for $J=0,1$ in
a  straight forward manner: The absorbtive parts are contained exclusively in
the functions $\bar{B}(\qsq,M_1^2,M_2^2)$, i.e.
\bea
{\rm Im} {1\ovr\i} \bar{B}(\qsq,M_1^2,M_2^2)&=&{1\ovr 16 \pi}
{\nu(\qsq,M_1^2,M_2^2)\ovr \qsq} \th(\qsq-(M_1+M_2)^2)
\qquad\mbox{with}
\\
\nu(\qsq,M_1^2,M_2^2)&=&\left[\qsq-(M_1+M_2)^2\right]^{1/2} 
\left[\qsq-(M_1-M_2)^2\right]^{1/2}
\; .
\eea
By simply rearranging contributions one observes that the result for $J=1$ can
be expressed in  terms of the on-shell vector form factor (\ref{f+kapi}), viz:
\beq
\rho_{V,s}^{(1)}=
{1\ovr 64 \pi^2}\left[ {\nu^3(\qsq,M_K^2,M_\pi^2)\ovr q^6}
\th(\qsq-(M_K+M_\pi)^2)+
(\pi \rightarrow \eta) \right] \cdot |f_+^{K\pi}(\qsq)|^2
\;.
\label{rho1final}
\eeq
Likewise, but with considerable more effort, it can be shown that the $J=0$
spectral function is given in terms of the on-shell scalar form factors 
(\ref{fnotkapi},~\ref{fnotkaet}), viz:
\beq
\rho_{V,s}^{(0)}=
{3(M_K^2-M_\pi^2)^2\ovr64\pi^2}
{\nu(\qsq,M_K^2,M_\pi^2)\ovr q^6}
\th(\qsq-(M_K+M_\pi)^2)\cdot |f_0^{K\pi}(\qsq)|^2+(\pi \rightarrow \eta)\; .
\label{rho0final}
\eeq
In deriving (\ref{rho1final},~\ref{rho0final}) we have used the GMO relation 
as well as the expansion
\beq
|f_{+,0}^{P,Q}|^2=
1+2\;{\rm Re}\Delta f_{+,0}^{P,Q}+O(p^4)
\eeq
which follows from $f_{+,0}^{P,Q}=1+{\rm Re}\Delta f_{+,0}^{P,Q}+
\i\,{\rm Im} \Delta f_{+,0}^{P,Q}$.
Eqns.~(\ref{rho1final},~\ref{rho0final}) are of course nothing else than the 
unitarity conditions for the absorbtive parts to the order we are interested in. 
Phenomenologically, the XPT expressions for the $J=1$ spectral function may
(at best) be used in the immediate threshold region only, since the prominent 
resonance $K^*$(892) is very close by. 
\end{itemize}

The physical content of our result can be further assessed by studying the 
asymptotic behavior $q^2\rightarrow\infty$ by means of the operator product 
expansion. Accordingly, the dispersion relations for
$\Pi^{(1+0)}_{V,s}(\qsq)$ and $q^2\Pi^{(0)}_{V,s}(\qsq)$ need at least one 
subtraction. In the chiral representation this fact is reflected by the
occurrence of contact terms, e.g. those in proportion to $H_1^{(0)}$ and
$C_{91}^{(0)}$. They depend on the particular way the ultraviolet divergences
have been regularized. These contact terms are seen to disappear from our
result upon either taking derivatives with respect to
$q^2$ or, in the case of the $J=1+0$ combination, by forming flavor breaking
differences of the flavor components $\Pi^{(1+0)}_{V,f}(\qsq)$ with
$f\in\{3,8,s\}$. Hence, the proper physical observables are
\bea
& &{d^n \over (d\qsq)^n}\;\Pi^{(1+0)}_{V,f}(\qsq), \quad 
{d^n \over (d\qsq)^n}\;\qsq \Pi^{(0)}_{V,f}(\qsq), \quad n=1,2,... \\
\label{Piderivatives}
{\rm and} \qquad
& &\Pi^{(1+0)}_{V,f_1}(\qsq)-\Pi^{(1+0)}_{V,f_2}(\qsq),\quad 
f_1,f_2\in\{3,8,s\}\; .
\label{Pidifference}
\eea
These functions depend on only two counterterm coupling constants of the 
$O(p^6)$ chiral Lagrangian, i.e. $P_V(\mu)$ and $Q_V(\mu)$. Since the 
constant $P_V(\mu)$ has already been determined to high accuracy
\cite{GoKa96}, we concentrate in the following on the flavor breaking 
differences (\ref{Pidifference}).


\subsection{A low energy theorem}

The general structure of the flavor breaking differences (\ref{Pidifference})
is
\bea
D^{(1+0)}_{f_1f_2}(\qsq) &\equiv&
\Pi^{(1+0)}_{V,f_1}(\qsq)-\Pi^{(1+0)}_{V,f_2}(\qsq)
\nonumber\\
&=& D_{f_1 f_2}^{\rm 1-loop}(\qsq)+D_{f_1 f_2}^{\rm 2-loop - cnt}(\qsq,\mu)+
D_{f_1 f_2}^{\rm cnt}(\qsq,\mu)\;.
\label{defDf1f2}
\eea
The first term in the last line of (\ref{defDf1f2}) denotes the total
one-loop contribution.
The second and third term represent the two-loop contribution which is
separated into finite loop- and the finite local $O(p^6)$ counterterm
pieces, respectively.
Explicit expressions for these terms can be obtained from 
(\ref{Sonetot},~\ref{Snottot}) and Ref.~\cite{GoKa95}. Note that the 
$O(p^4)$ counterterms $L_{10}^{(0)}$, $H_1^{(0)}$ drop out in such
differences.
Hence, the only terms containing free parameters are
\bea
D_{3 s}^{\rm cnt}(\qsq,\mu) &=& 4 {(M_K^2-M_\pi^2) \over F_0^2}\; Q_V(\mu)
\nonumber\\
D_{3 8}^{\rm cnt}(\qsq,\mu) &=& {16 \over 3} {(M_K^2-M_\pi^2) \over F_0^2}\;
Q_V(\mu)\; .
\label{Df1f2CT}
\eea
Consequently, there is one combination of vector-current correlators which is
free of any counterterm coupling constant, viz.
\beq
\Delta\Pi_V(\qsq)\equiv \Pi^{(1+0)}_{V,3}(\qsq)+3 \Pi^{(1+0)}_{V,8}(\qsq)
-4 \Pi^{(1+0)}_{V,s}(\qsq)=\Delta\Pi_V^{\rm 1-loop}(\qsq)+
\Delta\Pi_V^{\rm 2-loop}(\qsq)\; .
\label{LET}
\eeq
Eqn. (\ref{LET}) is a low energy theorem valid at the two-loop level. 
It shows that despite the many low energy constants occurring at this order of
the low energy expansion it is possible to give parameter free predictions 
for suitably chosen combinations of observables.

The RHS of (\ref{LET}) is readily evaluated numerically. At $\qsq=0$,
e.g., we obtain
\beq 
\Delta\Pi_V(0)=0.00513+0.00125
\;,
\eeq
where the 1-loop and 2-loop contributions have been displayed separately. We 
observe a correction relative to the leading order term of $\approx 25 \%$. 
This prediction for $\Delta\Pi_V(0)$ can be tested if the vector spectral
functions  $\rho_{V,f}^{(1+0)}$ are known to high accuracy. For instance, the
unsubtracted dispersion relation for $\Delta\Pi_V(\qsq)$ implies the chiral
sum rule
\beq
\Delta\Pi_V(0)=\int_{4 M_\pi^2}^\infty ds {\rho_{V,3}^{(1+0)}+
3 \rho_{V,8}^{(1+0)}-4 \rho_{V,s}^{(1+0)} \over s} \; .
\eeq
Saturating the spectral functions with narrow resonances for
$\rho(770)$, $\omega(782)$, $\phi(1020)$ and $K^*(892)$ gives the estimate
\beq
\Delta\Pi_V(0)\approx 0.039+0.085-0.121=0.003\; .
\label{NarrowRes}
\eeq
The high degree of cancellation present in (\ref{NarrowRes}) clearly calls for
the inclusion of finite width effects, higher resonances as well as
multiparticle contributions to the spectral strengths. However, currently
available data for hypercharge and strange components of the vector spectral
functions do not prove accurate enough to make this test conclusive.
We shall present a more detailed analysis involving the difference
$\rho_{V,3}^{(1+0)}-\rho_{V,s}^{(1+0)}$ in section V below.


\section{Inverse moment finite energy sum rules and determination of $Q_V$}


The two-loop representation of $\Pi_{V,s}^{(1,0)}$ given in the previous
sections can be combined with the isospin or hypercharge component of the
vector current correlators \cite{GoKa95,GoKa96} to form physically observable
quantities. In Ref.~\cite{GoKa96} the difference between correlators of
isospin and hypercharge currents was employed to estimate the finite 
counterterm $Q_V$. It has been pointed out later that isospin violating 
corrections have a rather large effect on the outcome of that analysis 
\cite{MW98}. The reason is that individual components get mostly
canceled such that the isospin breaking effect on the difference is enhanced.
Isospin violation affects primarily the hypercharge component and is hard
to be treated in a model-independent way.
The aim of this section is to present an analysis based on the difference of
isospin and strange components of the vector current which is less sensitive
to such corrections. Thus, we shall consider the flavor breaking difference 
\beq
\Pi(q^2)\equiv \Pi_{V,3}^{(1+0)}(q^2)-\Pi_{V,s}^{(1+0)}(q^2)
\;,
\label{defPi}
\eeq
which is free of kinematical singularities at the origin. The physical 
content of the XPT (chiral perturbation theory) representation of such
combinations is conveniently analyzed by means of inverse moment finite energy
sum rules (IMFESR) \cite{earlyFESR,Ste98}.
Unitarity and analyticity imply that the correlator (\ref{defPi}) satisfies
\beq
\oint_{C}ds\;{\Pi(s)w(s)\ovr s^{n+1}}=0
\;,
n=0,1,\ldots
\label{Cauchy}
\eeq
where C is the contour shown in Fig.~\ref{Figcontour} and $w(s)$ is a 
conveniently chosen weight function which is analytic inside this contour. 
In the following we shall consider polynomial weight functions with $w(0)=1$.
The contribution from the small circle can
be computed using the XPT representation of the correlator, i.e.
\beq
{1\ovr n!}{d^n\ovr (dq^2)^n}(\Pi w)(0)=
{1\ovr2\pi\i}\int_{|s|=s_0}ds\;{\Pi(s)w(s)\ovr s^{n+1}}+
\int_{s_{\rm th}}^{s_0}ds\;{\rh(s)w(s)\ovr s^{n+1}}
\;,\qquad n=0,1,\ldots
\label{IMFESR}
\eeq
with $\rho(s)$ being the spectral function of the correlator under
consideration, i.e.
\beq
\rh(s)={1\ovr \pi} {\rm Im} \Pi_{V,3}^{(1+0)}(s) \th(s-4 M_\pi^2)-
{1\ovr \pi} {\rm Im} \Pi_{V,s}^{(1+0)}(s) \th(s-(M_K+M_\pi)^2)
\;.
\label{rhodef}
\eeq

The first term on the right hand side of (\ref{IMFESR}) can reliably 
be evaluated using the operator product expansion for the correlator, 
provided the radius of the circle, $s_0$, is chosen large enough.
By contrast, the second term on the RHS of (\ref{IMFESR}) needs the difference
of hadronic isospin and strange vector spectral functions as input.
The only source of such hadronic spectral functions is provided by the tau
decay data \cite{ALEPHrho3,ALEPHrhos}; for this reason $s_0$ cannot be taken
much larger than the tau mass squared.

The weight function $w(s)$ needs to satisfy 3 more conditions to make
the calculation feasible.
First, $w(s)$ must be chosen in such a way that the perturbative series in
$\alpha_s$ of the leading OPE contributions is well controlled.
Second, the region close to the real axis, $s=s_0$, must be suppressed
since the OPE is not reliable in that region. Weight functions without this
suppression substantially violate local duality.
In practice, this suppression is highly welcome for reasons which relate to the
third condition:
As $s$ approaches the tau mass, data become more and more uncertain.
As a consequence, the integral over the hadronic spectral function is afflicted
with large error bars --- unless the weight function suppresses the
contributions from, say, $s\ge (1.4 {\rm GeV})^2$.
It is this point which makes the method of IMFESR superior to the conventional
inverse moment chiral sum rules \cite{GoKa96,DoGo94,GoKa97,GoKa98b}:
The high energy behavior of the weight function can be worse than what a once
subtracted dispersion relation would require. Nevertheless, as long as the
weight function exhibits one or two zeros at $s=s_0$, local duality is
extremely well satisfied, even at scales below $s=m_\tau^2$ \cite{MaltPLB440}.

In the following, we shall study the IMFESR (\ref{IMFESR}) for $n=0$ using the
weight-function
\beq
w(s)\equiv(1-x)^3\cdot(1+x+\hal x^2)
\;,\qquad
x\equiv{s\ovr s_0}
\;.
\label{wchrumm}
\eeq
This weight-function satisfies the conditions listed above.
The factor $(1-x)^3$ proves efficient in cutting off the region above
$2 {\rm GeV}^2$.

Next, we specify the input needed to evaluate the inverse moment finite energy
sum rule.
We shall consider in turn the XPT, the OPE, and the hadronic side of the IMFESR.


\subsection{Chiral constraints at $q^2=0$}

The correlator at $q^2=0$ is obtained by straight forward evaluation of the 
representation of $\Pi_{V,s}^{(1+0)}$ given in section III of this paper
and $\Pi_{V,3}^{(1+0)}$ as calculated in Ref.~\cite{GoKa95}. Some care has
to be taken in order to show the cancellation of singular terms in the limit
$q^2\rightarrow 0$. 
Employing input parameters $M_\pi=140$ MeV, $M_K=495$ MeV, $F_0=93$ MeV, and,
at renormalization scale $\mu=m_\rho$, $L_5^{(0)}=0.0014\pm 0.0005$ 
\cite{Eck94},
$L_9^{(0)}=0.00678\pm 0.00015$, $L_{10}^{(0)}=-0.00513\pm 0.00019$ 
\cite{Ste98} we obtain numerically
\beq
\Pi(0)=0.0053+0.0014+{4\ovr F_0^2}(M_K^2-M_\pi^2)Q_V(m_\rho^2)\;.
\label{PiChPTnum}
\eeq
The first and second term on the RHS of (\ref{PiChPTnum}) are finite one-
and two-loop terms, respectively. The counterterm coupling constant $Q_V$
has been estimated previously, with the result
\bea
Q_V(\mu=m_\rho) &=& (3.7\pm 2.0)\; 10^{-5} \qquad \cite{GoKa96}
\label{QVprevest1}
\\
Q_V(\mu=m_\rho) &=& (3.3\pm 0.4)\; 10^{-5} \qquad \cite{MW98}
\label{QVprevest2}
\;,
\eea
where the second number has been obtained with a phenomenological ansatz for
the 38-component of the vector spectral function.
Using the first of these values, the last term in (\ref{PiChPTnum}) yields 
numerically $0.0038\pm 0.0021$, i.e. at scale $\mu=m_\rho$ it is the dominant
effect of the two-loop contributions. 


\subsection{OPE of the correlator}

The contributions of dimension $D=2,4,6$ are known from previous work
\cite{JM95,CPS97,CFNP97}. The D=2 operators are just the flavor breaking 
quark mass corrections:
\beq
\Pi(Q^2)^{\rm D=2}={1\over Q^2}{3\over 4 \pi^2} \left\{
m_s^2 \Big[ 1+{7\over 3} a_s+19.9332 a_s^2+O(a_s^3)\Big]+
m_s m_u \Big[{2\over 3} a_s+O(a_s^2)\Big]+O(m_u^2)\right\},
\label{D2ope}
\eeq
where $Q^2=-q^2$ denotes the Euclidean momentum, and 
$a_s\equiv \alpha_s(Q^2)/\pi$. Since terms in proportion to $m_s m_u$ are 
further suppressed by $a_s$, keeping the $m_s^2$ term only provides an 
excellent approximation. 

Following the prescription of Ref.~\cite{LeDP92}, the integration of this 
function over the circular contour with radius 
$s_0=m_\tau^2$ is performed numerically. We use the 4-loop $\beta$- and 
$\gamma$-functions \cite{RVL97,Che97} for the running of mass and strong
coupling constant. Weighting the correlator with $w(s)$ given in 
(\ref{wchrumm}), we thus obtain
\beq
B^{\rm D=2} \equiv {1\ovr2\pi\i}\int_{|s|=s_0}ds\;{\Pi(s)w(s)\ovr s}=
{3\over 4\pi^2}{m_s^2(m_\tau^2)\over m_\tau^2}
\left[(1.634+0.303+0.177)\pm 0.088\right]
\;.
\label{BD2}
\eeq
The three terms in parenthesis represent the contributions at
$O(1),O(a_s),O(a_s^2)$ in the contour improved perturbation series.
The error reflects the uncertainty due to terms of order $a_s^3$, 
assuming a geometric growth of the series in $a_s$ at $Q^2=m_\tau^2$. The 
particular combination of moments appearing in the weight function $w(s)$ 
makes sure that even the resummed geometric series yields corrections
not larger than the size of the $a_s^3$ term. 

In evaluating Eqn.~(\ref{BD2}) the four-loop beta function is used,
but values of $a_s$ extracted are usually somewhat dependent on which order
of perturbation theory is used.
To get a handle on the uncertainty brought by the truncation in the
perturbative series, we mention that the bracket in Eqn.~(\ref{BD2})
is changed into $[1.668+0.316+0.190\pm0.096]$ if two-loop running is used.
In other words: $[2.174\pm0.096]$ instead of $[2.114\pm 0.088]$ is a
$3\%$ effect.

As a result, the total error is dominated by the uncertainty of present
evaluations of the  strange quark mass.
For a numerical evaluation we use $m_s(\mu=m_\tau)=(130\pm 25) {\rm MeV}$
\cite{JM95,CPS97,Mal99}, leading to $B^{\rm D=2}=(8.5\pm 3.3)\;10^{-4}$.
This corresponds to $\approx 16\%$ of the leading order one-loop term in
(\ref{PiChPTnum}).

The D=4 operators consist of two types, the quark mass corrections
and terms in proportion to the quark condensates. The leading quark mass 
correction is in proportion to $a_s^{-1} m_s^4$. Despite the $a_s^{-1}$
enhancement, the net contribution to the RHS of the sum rule (\ref{IMFESR})
is of the order $10^{-6}$, i.e. completely negligible compared to the D=2 
contributions found above. The condensate terms read 
\beq
\Pi(Q^2)_{\rm condensate}^{\rm D=4}=
{-1\over Q^4} {m_s\over \hat{m}} \langle \hat{m} \bar{u} u\rangle
\left[ r_c+a_s ({4\over 3}-r_c)+a_s^2 \left({59\over 6}-{13\ovr 3} r_c\right)
\right]. 
\eeq
The ratio $r_c\equiv\langle\bar{s}s\rangle/\langle\bar{u}u\rangle$ lies in
the range $0.7 \leq r_c \leq 1$ \cite{JM95,CPS97}. For numerical evaluation
we employ leading order XPT  expressions, i.e. $m_s/\hat{m}=25.9$
and $\langle \hat{m} \bar{u} u \rangle=-F_\pi^2 M_\pi^2/2$. 
In this manner, the contribution to the IMFESR under consideration 
is also found to be of the order $10^{-6}$. Hence, we can safely neglect
the contributions of $D\geq 4$ operators in the OPE.  


\subsection{Hadronic spectral function}

The isovector- and strange component of the vector spectral functions are 
constructed as follows:
i) close to threshold we employ the XPT two-loop expressions given in
\cite{GoKa95} and this paper. These results depend on the low energy constants
$L_5,L_9$. However, due to the smallness of the spectral functions in the
threshold region compared to the resonance region, uncertainties in these LECs
have no detectable effect on the IMFESR (\ref{IMFESR}) with n=0.
ii) above the threshold region and up to the tau mass we employ the data as
presented by the ALEPH collaboration \cite{ALEPHrho3,ALEPHrhos}. We discuss
this input as well as the corresponding evaluation of integrals over hadronic
spectral functions for isovector- and strange component in turn:
\begin{itemize}
\item[(a)]
isovector vector spectral function: the impressive set of ALEPH data
\cite{ALEPHrho3}
can be used over the whole range of integration. Data are accurate to 
$\approx 3$ \% for $s\leq 2 {\rm GeV}^2$. The region above this scale 
is however strongly suppressed when integrating with the weight function
(\ref{wchrumm}). Table 1 shows the hadronic integrals 
\beq
B^{{\rm had},J}_{V,f}(s_{\rm max},s_0)=\int_{s_{\rm th}}^{s_{\rm max}}ds\;
{\rh_f^J(s)w(s)\ovr s}
\label{Bhadtable}
\eeq
for various components $f$ of the vector spectral function and different 
upper integration ends $s_{\rm max}$. The parameter $s_0$ refers to the weight
function (\ref{wchrumm}) and is taken as large as possible, i.e.
$s_0=m_\tau^2\simeq(1.777~{\rm GeV})^2$.
The first line in Table~1 is split into three columns to illustrate the effect
of not extending the integration range in (\ref{Bhadtable}) to its upper end
$s_0$, but to 1 or 2~${\rm GeV}^2$ only (while keeping the weight function
(\ref{wchrumm}) with its zero at $s_0$). In the isovector case, the dominant
contribution to the hadronic integral stems from the region below
$\sim1~{\rm GeV}^2$, where the spectral function is still dominated by
individual resonances.
The error in $B^{\rm had}_{V,3}$ is predominantly due to the correlated
part of the error in the individual bins, i.e. mostly due to the uncertainty
of the $\tau$ branching ratios.
We are working in the isospin limit where the $J=0$ component 
of the isovector vector spectral function vanishes.
\item[(b)]
strange component of the vector spectral function, $J=1+0$: in a recent 
publication 
\cite{ALEPHrhos} the ALEPH collaboration has presented results for the 
strange component of vector plus axial-vector spectral functions obtained from 
tau decay data. Unlike in the isovector case, a full separation of vector and
axial-vector components was not achieved. However, as we shall argue below, 
the information is sufficient to evaluate the difference of hadronic integrals
we are seeking to an accuracy of $\approx 17$ \%. To this end we analyze the
data according to the hadronic state $X$ in the decay
$\tau\rightarrow\nu_\tau+X$:
\begin{itemize}
\item
$X=K\pi$: the decay to the $K\pi$ final state is a pure vector current 
transition. The main feature of this component is the $K^*$(892) resonance, 
but there is a minor contribution associated with the interference of
$K^*$(1410). The resulting integral $B^{\rm had}_{K\pi}$
is given in the third line of Table 1. Again, the error is primarily due
to the uncertainty in the overall normalization.
\item
$X=K 2\pi$: the final state with two pions is due to both, vector- and 
axial-vector transitions. As detailed in \cite{ALEPHrhos} the corresponding 
spectral function can be understood in terms of the two axial-vector 
resonances $K_1$(1270), $K_1$(1400) and the vector $K^*$(1410). The mass 
resolution and statistics are not sufficient to separate the $K_1$(1400)
and $K^*$(1410). We follow \cite{ALEPHrhos} and use an effective resonance
instead, averaging the parameters of the two states. The vector part is then
estimated by fitting $\rho_{v,s}^{(1)}+\rho_{a,s}^{(1)}$ to two resonances 
with masses $m_1=1.270$ GeV, $m_2=1.405$ GeV. Switching off the axial 
$K_1$(1270) yields an estimate for the contribution from the resonances at 
$s=(1.4\; {\rm GeV})^2$. Half of the remaining spectral strength is assumed to 
be vector, with an estimated error of $\approx 50$ \%. This procedure is 
roughly consistent with the vector- axial-vector separation of branching ratios
shown in Table 7 of Ref.~\cite{ALEPHrhos}. The contribution to the hadronic
integral $B^{\rm had}_{K 2\pi}$ is roughly 7 \% of the $K\pi$ component,
see Table~1.
\item
$X=K\eta$ and $K 3\pi$: this component has not been separated into vector- and 
axial-vector contribution. However, the spectral strength associated with 
these final states is relevant only above $s\approx (1.3\; {\rm GeV})^2$. 
Due to the suppression of this energy region by the weight function, the
net contribution to the hadronic integral is small.
We assign half of this component to the vector current and employ the rest 
of the spectral strength as an estimate for the error.
\item
$X=Kn\pi$ for $n\geq 4$: these components are small and non-zero only above
$s=2\;{\rm GeV}^2$. For this reason they can safely be neglected. 
\end{itemize}
Again, the different columns in Table~1 illustrate the effect of not extending
the integration range in (\ref{Bhadtable}) to its upper end $s_0$, but to
1 or 2~${\rm GeV}^2$ only (while keeping the weight function (\ref{wchrumm})
with its zero at $s_0$). The contributions to the strange vector spectral
function are not dominated by the low-energy part as impressively as the
non-strange part, but still the integrals are almost saturated at
$\sim1~{\rm GeV}^2$.
\item[(c)]
strange component of the vector spectral function, $J=0$:
the main contribution to the exclusive $J=0$ component is provided by 
production of the scalar resonance $K_0^*$(1430), which decays almost 
exclusively to the $K\pi$ final state \cite{ALEPHrhos}. In order to obtain 
an estimate for this component we use an Omn\`es representation as given in 
Ref.~\cite{CFNP97}. Employing also input parameters as detailed in
\cite{CFNP97} yields the integrals as shown in Table 1. The error is an 
educated guess of the various uncertainties entering the Omn\`es
representation. This effect is already accounted for in the $K\pi$ component
and should not be included when summing the strange contributions to
$B^{\rm had}_{V,s}$.
\end{itemize}

The hadronic integrals summed over the strange components of the vector 
spectral functions is given in the third to last line of Table~1. 
Subtracting this result from the hadronic integral over the isovector 
component and adding errors in quadrature we finally obtain
\beq
B^{\rm had}_{V,3}-B^{\rm had}_{V,s}=0.0087 \pm 0.0013 \;.
\label{Bhadnum}
\eeq
Individual flavor components thus have canceled to $\approx~60$~\%. The 
error stems mostly from the error on the $K\pi$-component of the strangeness 
carrying vector spectral function, but the error on the $K2\pi$-component
is non-negligible as well.

The final step is plugging the results of Eqns.~(\ref{PiChPTnum},~\ref{BD2},
\ref{Bhadnum}) into the IMFESR (\ref{IMFESR}) for $n=0$, which yields
\beq
Q_V(\mu=m_\rho)=(2.8 \pm 1.3)\cdot 10^{-5}\; .
\label{QVresult}
\eeq
This result is consistent with the determinations (\ref{QVprevest1},
\ref{QVprevest2}), but with the advantage of being both, relatively precise
and free of model assumptions.

The variation of $Q_V$ with renormalization scale $\mu$ is known 
\cite{GoKa95}: 
\beq
Q_V(\mu)-Q_V(\mu_0)=-{3\over32\pi^2}
\left(L_9^{(0)}(M_\rho)+L_{10}^{(0)}(M_\rho)\right)
\ln{\mu^2\over\mu_0^2}
\;.
\label{QVscale}
\eeq
$Q_V$ is thus strongly varying with scale. Explicitly we obtain e.g.
$Q_V({\rm 500 MeV})=(4.2\pm 1.4) 10^{-5}$ and 
$Q_V({\rm 1 GeV})=(2.0\pm 1.4) 10^{-5}$. 
However, the method employed here deals with RG-invariant quantities,
and the variation with scale does not lead to any further ambiguities
in the determination of $Q_V$. We also note that the coupling $Q_V$ is
the same in both, the $\bar{\lambda}$-subtraction scheme used here and 
the $\bar{{\rm MS}}$-scheme \cite{GoKa98a}.

\begin{figure}[p]
\begin{center}
\epsfxsize=10cm
\epsffile{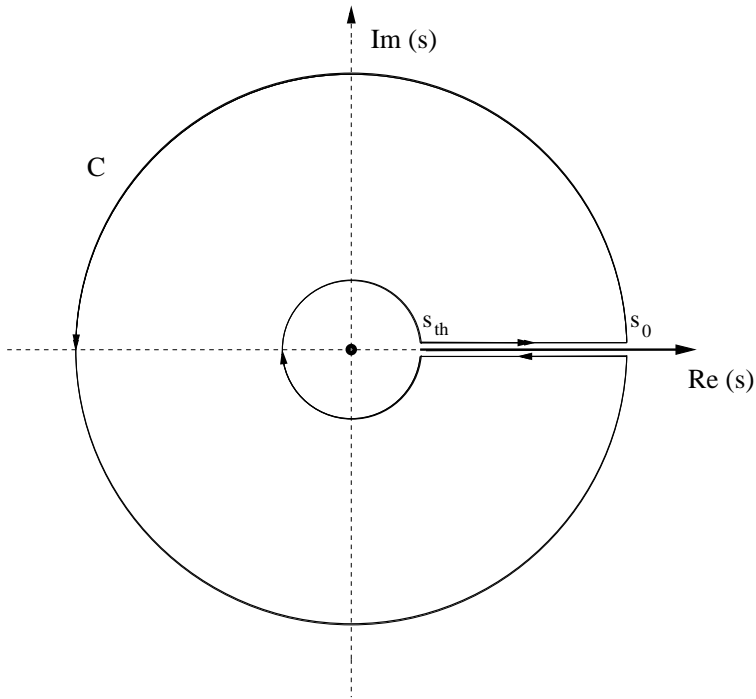}
\vspace{0.5cm}
\caption[]{\sl Integration contour for the inverse moment finite energy sum
rule (\ref{Cauchy}). The cut on the real axis starts at $s_{\rm th}$ and
extends to infinity. The dot at the origin indicates possible poles in the
case of inverse moment finite energy sum rules.}
\label{Figcontour}
\end{center}
\end{figure}

In our determination, $s_0$ was chosen as large as possible by having it equal
to $m_\tau^2$. Of course one may ask whether the total result is stable under a
variation of the radius of the outer circle in Fig.~\ref{Figcontour}.
Since there are no data from $\tau$-decays with $s>m_\tau^2$, the only
alternative is a radius which is smaller than our choice $s_0=m_\tau^2$.
Concerning the contribution from the hadronic spectral integral, there is a
clear numerical answer to this question. Repeating our calculation, now
replacing [both, in the upper boundary of the integral (\ref{Bhadtable}) and
in the weight function (\ref{wchrumm})] $s_0=(1.777~{\rm GeV})^2$ by
$(1.555~{\rm GeV})^2$ and $(1.333~{\rm GeV})^2$, we find the values collected
in Table 2. As one can see, the total hadronic spectral integral ($3-s$
component) is extremely stable under this shift.
On the other hand any numerical statement about the shift in the contribution
from the outer circle is somewhat questionable. 
Lowering the radius in the integral (\ref{BD2}) to $|s|=s_1=(1.555~{\rm GeV})^2$
results in the bracket in the r.h.s. being replaced by
$[1.570+0.302+0.180\pm0.090]$. Going further down to a radius as small as
$|s|=s_2=(1.333~{\rm GeV})^2$ gives a bracket which takes the value
$[1.474+0.296+0.181\pm0.093]$.
The corresponding shift in $B^{D=2}$ is small compared to the error brought
by the uncertainty in present evaluations of the strange quark mass.
Hence, the contribution from the outer circle seems to be insensitive to the
precise value of the radius too.
The real problem, however, is that the expansion in contributions at
different orders in the OPE breaks down at $s$-values substantially below
$4 {\rm GeV}^2$. In this respect choosing $s_0=m_\tau^2$ amounts to stretching
the validity of the perturbative expansion to the very end. All one can say is
that our result is stable under {\em small\/} variations of the outer radius in
Fig.~\ref{Figcontour}.

Regarding the error, it is clear that a more accurate determination of both the
strange and the non-strange $\tau$ branching ratios as well as a more accurate
measurement of the $K\pi$-component of the vector spectral function below
$s=(1.4\ {\rm GeV})^2$, in particular in the $K^*$(892) region, will transform
into a more precise determination $Q_V$. As for the $K2\pi$ component,
a full separation into V- and A- part would further diminish the error. 

Our final value for $Q_V$ supports the claim by Maltman and Wolfe \cite{MW98}
that isospin breaking effects tend to reduce the estimate (\ref{QVprevest1}).


\section{Conclusions}


We have calculated the two-point function of strangeness carrying vector 
currents to two-loop standard chiral perturbation theory, in the isospin 
limit.
Our fully renormalized finite result passes several consistency checks:
The sum of all contributions does not depend on the renormalization scale $\mu$.
In the $SU(3)$ limit of equal quark masses it reduces to the known results for
the two-point functions of isospin and hypercharge currents. The $SU(3)$
breaking part satisfies the constraint imposed by the Ademollo Gatto theorem.
Furthermore, kinematical singularities at zero momentum transfer present in the
spin 0 and spin 1 components are shown to be absent if a convenient and
well-known decomposition is chosen. Last but not least, the imaginary parts of
the correlators for spin 0 and 1 obey the unitarity-conditions along the
two-particle cut implying that they can be expressed in terms of the
vector and scalar meson form-factors.

The calculated correlators $\Pi_{V,s}^{(J)}$ for J=0,1 depend on several low
energy constants (LEC). Two of these constants are related to contact terms.
In physical observables, which do not depend on contact terms, only two low
energy constants of the $O(p^6)$ chiral Lagrangian remain, i.e. $P_V$ and $Q_V$.
We have shown that at $O(p^6)$ in the low energy expansion the combination
$(\Pi_{V,3}^{(1+0)}+3 \Pi_{V,8}^{(1+0)}-4 \Pi_{V,s}^{(1+0)})(q^2)$ is free of
any unknown LEC and consequently predicted in terms of pseudoscalar meson
masses and decay constants only.
This low energy theorem (LET) can be tested once the vector spectral functions
$\rho_{V,3}$, $\rho_{V,8}$ and $\rho_{V,s}$ are known to high accuracy. 
Presently, data are not accurate enough to make such a test conclusive, but
the situation might change within a couple of years. 

Finally we have presented a determination of the low energy constant $Q_V$
based on a inverse moment finite energy sum rule. The sum rule employs the 
difference of isospin and strange components of vector spectral functions 
and is therefore expected to be less sensitive to isospin breaking effects 
than previous attempts. Our analysis is based on the recently released ALEPH
data of $\tau$-decays into strangeness carrying final states. Presently, the
error bars on the data are still too large to make the uncertainty of our
first-principle based $Q_V$ determination smaller than the statistical
error of previous model-based calculations. However, the fact that the main
uncertainty stems from the dominant decay channel means that our hope for
more precise experimental data might be realistic. The very same experimental
effort would also be welcome in view of the low energy theorem discussed above.


\vfill
\subsection*{Acknowledgments}
\vspace{-10pt}
\noindent
We are grateful to G.~Colangelo, J.~Gasser and K.~Maltman for illuminating
discussions and to S.~Chen for providing us with data on the strange component
of the vector spectral function.
This work is supported in part by U.S. Department of Energy grant
DE-FG03-96ER40956 and the EEC-TMR Program, Contract No CT98-0169.
Both authors acknowledge support from the Swiss National Science Foundation.

\clearpage


\section*{Appendix: Survey of One-loop Integrals}


Here we give a survey of the integrals occurring in the computation of the
formfactor and the desired correlator.
We shall keep $d$ arbitrary and isolate the divergences near $d\to4$.




The simplest quantity is the scalar integral
\beq
A(m^2)=\int{d^d\ell\ovr(2\pi)^d}\,{1\ovr\ell^2-m^2}
\label{Adef}
\eeq
which, by standard means, is evaluated to give
\beq
A(m^2)={-\i\ovr(4\pi)^{d/2}}\,\Gamma(1-{d\ovr2})\,(m^2)^{(d/2-1)}
\;.
\label{Aeval}
\eeq
From this expression it is found to split
\bea
A(m^2)&=&-2\i\,m^2\,\mu^{(d-4)}\;\lab+\Afin(m^2)
\qquad\mbox{with}
\label{Asplit}
\\
\lab&=&{1\ovr16\pi^2}\,\Big({1\ovr d-4}-\hal(\log(4\pi)-\ga+1)\Big)
\label{labdef}
\\
\Afin(m^2)&=&\mu^{(d-4)}\,\Big(
-{\i m^2\ovr 16\pi^2}\log({m^2\ovr\mu^2})
+\ldots
\Big)
\label{Afin}
\eea
where (\ref{Afin}) is the part of $A$ which stays finite under $d\to4$ in this
so-called $\lab$-scheme.
Each of the contributions in (\ref{Asplit}) depends on the artificial scale
$\mu$ and the prefactor $\mu^{(d-4)}$ is introduced in
order to give them the appropriate mass dimension in $d$ space-time dimensions.



We shall define
\bea
B(q^2,\mo,\mt)&=&\int{d^d\ell\ovr(2\pi)^d}\,
{1\ovr(\ell^2-\mo)\,((\ell-q)^2-\mt)}
\label{Bdef}
\\
\Bmu(q,\mo,\mt)&=&\int{d^d\ell\ovr(2\pi)^d}\,
{q_\mu\ovr(\ell^2-\mo)\,((\ell-q)^2-\mt)}
\label{Bmudef}
\\
\Bmunu(q,\mo,\mt)&=&\int{d^d\ell\ovr(2\pi)^d}\,
{q_\mu q_\nu\ovr(\ell^2-\mo)\,((\ell-q)^2-\mt)}
\label{Bmunudef}
\eea
as well as the finite quantity
\beq
\Bbar(q^2,\mo,\mt)=B(q^2,\mo,\mt)-B(0,\mo,\mt)
\label{Bbar}
\eeq
which obviously satisfies $\Bbar(0,\mo,\mt)=0$.
Then the exact relations (in $d$ dimensions)
\bea
B(0,m^2,m^2)&=&\,{(d-2)\ovr2}\;{A(m^2)\ovr m^2}
\label{Bzersame}
\\
B(0,\mo,\mt)&=&{A(\mo)-A(\mt)\ovr\mo-\mt}
\label{Bzerdist}
\eea
follow by taking a derivative in (\ref{Adef},~\ref{Aeval}) and by partial
fraction decomposition in (\ref{Bdef}).
Invoking covariance, $\Bmu$ and $\Bmunu$ in Eqns.~(\ref{Bmudef},~\ref{Bmunudef})
may be decomposed as
\bea
\Bmu(q,\mo,\mt)&=&\qmu B_1(\qsq,\mo,\mt)
\label{Bmudec}
\\
\Bmunu(q,\mo,\mt)&=&\qmu\qnu B_{21}(\qsq,\mo,\mt)+\gmunu B_{22}(\qsq,\mo,\mt)
\;.
\label{Bmunudec}
\eea
Considering $\qmu\Bmu$ one finds
\bea
B_1(\qsq,\mo,\mt)
&=&
\hal\,(1+{\mo-\mt\ovr\qsq})\,B(\qsq,\mo,\mt)-{(A(\mo)-A(\mt))\ovr2\qsq}
\label{Boneorig}
\\
&=&
\hal\,(1+{\mo-\mt\ovr\qsq})\,\Bbar(\qsq,\mo,\mt)+{(A(\mo)-A(\mt))\ovr2(\mo-\mt)}
\label{Bonedeco}
\eea
where in (\ref{Bonedeco}) one would have to use Eqns.~(\ref{Bzersame},
\ref{Bzerdist}) for performing the equal mass limit.
Considering $\qmu\Bmunu$ and $\gmunu\Bmunu$ one finds
\bea
(d-1)\,B_{21}
&=&
{d\ovr2}\,(1+{\mo-\mt\ovr\qsq})\,B_1
-\,{\mo\ovr\qsq}\,B
+{d-2\ovr2\qsq}\,A(\mt)
\label{Btwooneorig}
\\
&=&
{d(\qsq+\mo-\mt)^2-4\qsq\mo\ovr4\qsq}\;\Bbar
\nonumber
\\
&{}&
-\,{(4-d)\mo-d(\qsq-\mt)\ovr4(\mo-\mt)\qsq}\,A(\mo)
-{d(\qsq-\mo)-(4-d)\mt\ovr4(\mo-\mt)\qsq}\,A(\mt)
\label{Btwoonedeco}
\\
(d-1)\,B_{22}
&=&
-\,\hal(\qsq+\mo-\mt)\,B_1
+\,\mo\,B
-\hal\,A(\mt)
\label{Btwotwoorig}
\\
&=&
-\,{q^4-2\qsq(\mo+\mt)+(\mo-\mt)^2\ovr4\qsq}\;\Bbar
\nonumber
\\
&{}&
-\,{\qsq-3\mo-\mt\ovr4(\mo-\mt)}\,A(\mo)
+{\qsq-\mo-3\mt\ovr4(\mo-\mt)}\,A(\mt)
\label{Btwotwodeco}
\eea
where in (\ref{Btwoonedeco}) and (\ref{Btwotwodeco}) one would have to use Eqns.
(\ref{Bzersame},~\ref{Bzerdist}) for performing the equal mass limit.
A combination frequently occurring in loop integrals is
\bea
\Tmunu(q,\mo,\mt)
&=&
4\Bmunu-2\qmu\Bnu-2\qnu\Bmu+\qmu\qnu B
\label{Tmunudef}
\\
&=&
\gmunu4B_{22}+\qmu\qnu(4B_{21}-4B_{1}+B)
\label{Tmunuorig}
\\
&=&
P^{(0)}_{\mu\nu}\cdot
\Big(4B_{22}+(4B_{21}-4B_{1}+B)\,\qsq\Big)
+
P^{(1)}_{\mu\nu}\cdot
\Big(4B_{22}\Big)
\label{Tmunudeco}
\eea
where, in the last line, we have introduced the projection operators
\beq
P^{(0)}_{\mu\nu}={\qmu\qnu\ovr\qsq}
\quad\mbox{and}\quad
P^{(1)}_{\mu\nu}=\gmunu-{\qmu\qnu\ovr\qsq}
\;.
\label{projectors}
\eeq
Denoting $T^{(0)}=4B_{22}+(4B_{21}-4B_{1}+B)\,\qsq$ and $T^{(1)}=4B_{22}$ the
spin-0 and spin-1 components, respectively, and using the decompositions
(\ref{Bonedeco},~\ref{Btwoonedeco},~\ref{Btwotwodeco}) along with
(\ref{Bbar},~\ref{Bzersame},~\ref{Bzerdist}) these two pieces are found to
split up into divergent and finite parts
\bea
T^{(0)}_{\rm di\!v}(\qsq,\mo,\mt)
&=&
-2\i\,(\mo+\mt)\,\mu^{(d-4)}\;\lab
\label{Tnotdiv}
\\
T^{(0)}_{\rm fin}\,(\qsq,\mo,\mt)
&=&
{(\mo-\mt)^2\ovr\qsq}\;\Bbar(\qsq,\mo,\mt)+\Afin(\mo)+\Afin(\mt)
\label{Tnotfin}
\\
T^{(1)}_{\rm di\!v}(\qsq,\mo,\mt)
&=&
{2\i\ovr3}\;(\qsq-3(\mo+\mt))\,\mu^{(d-4)}\;\lab
\label{Tonediv}
\\
T^{(1)}_{\rm fin}(\qsq,\mo,\mt)
&=&
-{1\ovr3}
\Big\{
{q^4-2\qsq(\mo+\mt)+(\mo-\mt)^2\ovr\qsq}\;\Bbar(\qsq,\mo,\mt)
\nonumber
\\
&{}&
\qquad
+\,{\qsq-3\mo-\mt\ovr\mo-\mt}\,\Afin(\mo)
-\,{\qsq-\mo-3\mt\ovr\mo-\mt}\,\Afin(\mt)
\nonumber
\\
&{}&
\qquad
+{\i\ovr24\pi^2}(\qsq-3(\mo+\mt))
\,\Big\}
\;,
\label{Tonefin}
\eea
where terms contributing to $O(d-4)$ have been dropped and
where the equal-mass limit is trivial for every quantity but
$T^{(1)}_{\rm fin}$, for which it takes the form
\beq
T^{(1)}_{\rm fin}(\qsq,\msq,\msq)=
-{1\ovr3}\Big\{
(\qsq-4\msq)\;\Bbar(\qsq,\msq,\msq)
+(\qsq-6\msq){\Afin(\msq)\ovr\msq}
-{\i\qsq\ovr48\pi^2}
\Big\}
\;.
\label{Tonefinsame}
\eeq


\bigskip
\vspace*{1cm}

\bdm
\mbox{TABLES}
\edm

\bigskip

\begin{tabular}{lccc}
\hline
\hline
$s_{\rm max}(\leq s_0):$ &
$1{\rm GeV}^2$ & $2{\rm GeV}^2$ & $s_0=(1.777~{\rm GeV})^2$\\
\hline
3-component, J=(1)=(1+0) &
$0.0240 \pm 0.0005$ & $0.0254 \pm 0.0005$ & $0.0257 \pm 0.0005$\\
\hline
$K\pi$-component &
$0.0146 \pm 0.0009$ & $0.0155 \pm 0.0010$ & $0.0155 \pm 0.0010$\\
$K 2\pi$-component  &
$0.0000 \pm 0.0002$ & $0.0008 \pm 0.0004$ & $0.0011 \pm 0.0006$\\
$K\eta$- and $K3\pi$-component &
$0.0000 \pm 0.0002$ & $0.0003 \pm 0.0003$ & $0.0004 \pm 0.0004$\\
\hline
sum s-component, J=(1+0) &
$0.0146 \pm 0.0009$ & $0.0166 \pm 0.0011$ & $0.0170 \pm 0.0012$\\
\hline
3-s, J=(1+0) & & &
$0.0087 \pm 0.0013$\\
\hline
s-component, J=(0) & & &
$0.0011 \pm 0.0004$\\
\hline
\hline
\end{tabular}
\vspace*{0.3cm}
\newline
Table~I:~{\sl Hadronic integrals $B^{\rm had,J}_{V,f}(s_{\rm max},s_0)$ as
defined in (\ref{Bhadtable}) for various components $f$ of the total vector
spectral function $\rho^{\rm J}_{V,f}$ as defined in Eqn.~(\ref{rhodef}).
Weight function is (\ref{wchrumm}) with $s_0=m_\tau^2=(1.777~{\rm GeV})^2$.}

\vspace*{1cm}

\begin{tabular}{lcc}
\hline
\hline
{}& 
$s_{\rm max}=s_2=(1.333{\rm GeV})^2$ &
$s_{\rm max}=s_1=(1.555{\rm GeV})^2$\\
\hline
3-component, J=(1)=(1+0) &
$0.0160 \pm 0.0004$ &
$0.0213 \pm 0.0005$\\
\hline
$K\pi$-component &
$0.0073 \pm 0.0005$ &
$0.0119 \pm 0.0008$\\
$K 2\pi$-component &
$0.0000 \pm 0.0001$ &
$0.0003 \pm 0.0003$\\
$K\eta$- and $K3\pi$-component &
$0.0000 \pm 0.0001$ &
$0.0001 \pm 0.0002$\\
\hline
sum s-component, J=(1+0) &
$0.0073 \pm 0.0005$ &
$0.0123 \pm 0.0009$\\
\hline
3-s, J=(1+0) &
$0.0087 \pm 0.0006$ &
$0.0090 \pm 0.0010$\\
\hline
s-component, J=(0) &
$0.0003 \pm 0.0001$ &
$0.0007 \pm 0.0003$\\
\hline
\hline
\end{tabular}
\vspace*{0.3cm}
\newline
Table~II:~{\sl Hadronic integrals $B^{\rm had,J}_{V,f}(s_2,s_2)$ as defined in
(\ref{Bhadtable}) for various components $f$ of the total vector spectral
function $\rho^{\rm J}_{V,f}$ as defined in Eqn.~(\ref{rhodef}). Weight
function is (\ref{wchrumm}) with $s_0\rightarrow s_1=(1.555~{\rm GeV})^2$
and $s_0\rightarrow s_2=(1.333~{\rm GeV})^2$, respectively. Either one
of the two columns corresponds to the rightmost column in Table I.}




\end{document}